\documentclass[conference]{IEEEtran}

\usepackage{cite}
\usepackage{easybmat}
\usepackage{stfloats}
\usepackage{amsmath,amssymb,amsfonts}
\usepackage{graphicx}
\usepackage{textcomp}
\usepackage[linesnumbered,lined,boxed,commentsnumbered, ruled]{algorithm2e}
\usepackage{epstopdf}
\usepackage{balance}
\usepackage{caption}
\usepackage{subcaption}
\usepackage{algorithmic}
\usepackage{xcolor}
\usepackage[letterpaper, left=0.7in, right=0.7in, bottom=1in, top=0.7in]{geometry}

\newcommand{\mrm}[1]{\ensuremath{\mathrm{#1}}}
\renewcommand{\vec}{\ensuremath{\mrm{vec}}}
\newcommand{\diag}{\ensuremath{\mrm{diag}}}
\newcommand{\bs}[1]{\ensuremath{\boldsymbol{#1}}}
\newcommand{\comment}[1]{}

\renewcommand{\H}{\boldsymbol{H}}
\newcommand{\RISA}{ARIS}
\newcommand{\RISAA}{FRIS}

\begin{document}

\title{Channel Orthogonalization with Reconfigurable Surfaces}
\author{Juan Vidal Alegr\'{i}a\IEEEauthorrefmark{1}, Fredrik Rusek\IEEEauthorrefmark{1}\IEEEauthorrefmark{2} \\
\IEEEauthorblockA{\IEEEauthorrefmark{1}Department of Electrical and Information Technology, Lund University, Lund, Sweden\\} 
\IEEEauthorblockA{\IEEEauthorrefmark{2}Sony Research Center, Lund, Sweden\\} 
\{juan.vidal\_alegria@eit.lth.se, fredrik.rusek@eit.lth.se\}}

\maketitle

\begin{abstract}
Orthogonal multi-user multiple-input multiple-output (MU-MIMO) channels allow for optimum performance with simplified precoding/equalization, and they achieve maximum multiplexing gain which is shared fairly among users. Reconfigurable intelligent surface (RIS) constitutes a promising cost-efficient solution to improve the wireless channel, since they consist of passive reflecting elements able to adjust the phases of the incoming waves. However, it is still widely unclear how these surfaces can improve spatial-multiplexing. In fact, the common RIS model cannot achieve perfect orthogonalization of MU-MIMO channels with a reasonable number of elements. Furthermore, efficient channel estimation algorithms for RIS, which are key for taking advantage of its benefits, are still a matter of research. We study two types of reconfigurable surfaces (RSs), namely amplitude-reconfigurable intelligent surface (ARIS) and fully-reconfigurable intelligent surface (FRIS), with extended capabilities over RIS. We show how these RSs allow for perfect channel orthogonalization, and, by minimizing the applied power, we show that they can potentially be implemented without the need of amplification. We also present an efficient channel estimation method for each of them that allows the base station (BS) to select the desired propagation channel.
\end{abstract}

\section{Introduction}\label{section:intro}
Multi-user multiple-input multiple-output (MU-MIMO) \cite{jindal}, has become a standard solution for base station (BS) implementation in 5G networks. After the first commercial deployments of Massive MIMO \cite{marzetta,emil_next}, its large scale counterpart, MU-MIMO is now a mature technology that allows multiplexing user equipments (UEs) in the spatial domain. However, the ability to exploit multiplexing gains with MU-MIMO depends on the conditions of the wireless propagation channels.

Reconfigurable intelligent surface (RIS) has emerged as a promising enabling technology towards future generation networks \cite{RIS,emil_next,HMIMOS}. Also known as intelligent reflective surface (IRS), this technology works as a passive reflector which can adjust the propagation environment in a power and cost-efficient manner. The reflected waves at the RIS can be redirected to create constructive interference and increase the received signal, leading to impressive gains in energy efficiency \cite{pw_scale,en_eff}.

Previous work has also considered RIS for improving spatial multiplexing in MIMO settings. For example, \cite{rank_impr} uses RIS for improving the rank of a single-user MIMO channel. RIS has also been considered for maximizing the user rates in different settings \cite{sum_rate,rate}. However, most of the previous results rely on the availability of channel state information at the BS, while channel estimation in RIS scenarios becomes extremely challenging \cite{HMIMOS} due to its limited capabilities and large number of elements. 

In this work, we study two alternatives to RIS, namely amplitude-reconfigurable surface (ARIS) and fully-reconfigurable surface (FRIS), for orthogonalizing MU-MIMO channels. To the best of our knowledge, the available research has not considered the fundamental problem of using reconfigurable surfaces (RS) for obtaining orthogonal MIMO channels, while this is of essential interest since it leads to full-multiplexing gain with fair user sharing (same channel power per UE) \cite{mimo}. Furthermore, for these channels, optimal processing at the base station is achieved by simple maximum ratio combining/transmission (MRC/MRT). We present a channel estimation method for each RS that allows the BS to select its own channel by computing and sending the RS configuration with a reduced number of pilots. We also show that these RSs can be realized without the need for amplification by minimizing the required power.

The rest of the paper is organized as follows. Section~\ref{section:model} describes the system model, and defines the different RSs capabilities. Section~\ref{sec:orth} describes how to achieve perfectly orthogonal channels with RSs. In Section~\ref{sec:ch}, we present the channel estimation processes for configuring the RSs. Section~\ref{sec:min} the power minimization setting. Section~\ref{section:num_res} gives numerical results. The paper is concluded in Section~\ref{section:conc}.

\section{System model}\label{section:model}
Let us consider an uplink MU-MIMO scenario where $K$ UEs are transmitting to an $M$-antenna BS, with $M>K$, through a narrow-band channel with the aid of an $N$-element RS. The $M\times 1$ received complex vector, $\boldsymbol{y}$, can be expressed as
\begin{equation}\label{eq:ul_model}
\bs{y} = \bs{H}\bs{s} + \bs{n},
\end{equation}
where $\bs{H}$ is the $M\times K$ channel matrix, $\boldsymbol{s}$ is the $K \times 1$ vector of symbols transmitted by the UEs, with $\mathbb{E}(|s_k|^2)=E_s$ $\forall k$, and $\boldsymbol{n}\sim \mathcal{CN}(\boldsymbol{0}_{M\times 1}, N_0\mathbf{I}_M)$ is the noise vector. Considering that there exists a direct channel, as well as a reflected channel through the RS, we can express the channel matrix as
\begin{equation}\label{eq:channel}
\H = \H_0+\H_1 \bs{\Theta} \H_2,
\end{equation}
where $\H_0$ corresponds to the $M\times K$ direct channel between the BS and the UEs, $\H_1$ and $\H_2$ correspond to the $M\times N$ channel between the BS and the RS and the $N \times K$ channel between the RS and the UEs, respectively, and $\bs{\Theta}$ is the reflection matrix applied at the RS. 

In the literature, it is common to restrict the RS to have phase shifting capabilities. This corresponds to the widely known concept of RIS where
\begin{equation}\label{eq:mod_RIS}
    \bs{\Theta}_\mrm{RIS} = \diag\left(\exp(j\phi_1),\dots,\exp(j\phi_N)\right).
\end{equation}
 In this paper, however, we propose two RS technologies where said restriction is relaxed, and we compare them in the task of orthogonalizing the channel matrix.
 
Let us consider an RS, here referred to as \RISA, whose elements can also adjust amplitude. The corresponding reflection matrix is then defined by
 \begin{equation}
    \bs{\Theta}_\mrm{\RISA} = \diag\left(\alpha_1,\dots,\alpha_N\right),\;\; \alpha_i\in \mathbb{C} \;\; \forall i.
\end{equation}
Note that the restriction of having each $\alpha_i$ of amplitude $1$ is here relaxed. The idea of adding amplification to a RIS has already been considered in the literature, and some of the hardware implications to realize these systems are given in \cite{ARIS_basar,ARIS_larsson}. However, one of our goals is to restrict the power of these surfaces so that they can still be implemented without the need for active amplification.

We also consider an RS, here referred to as \RISAA, whose reflection matrix is a complete matrix. Thus, we have 
\begin{equation}
    \bs{\Theta}_\mrm{\RISAA}\in \mathbb{C}^{N\times N}.
\end{equation}
In this work, we will not elaborate on the challenges of realizing such a RS. However, we can think of architectures based on vector modulators such that the matrix multiplication can be performed by an analog combiner as in \cite{VM_nir}, although, if future technology allows it, fully-digital implementations would be desirable so that processing is done per sub-carrier.

\section{Channel orthogonalization}\label{sec:orth}
The main goal of employing RSs is to adjust the propagation channel to make it more beneficial in some metric, e.g., array gain, channel capacity, multiplexing gain, etc. Within the considered framework, orthogonal channels\footnote{A more accurate term would be unitary channels due to the complex nature of the channel matrix.} are channels whose columns are constructed from unitary matrices, i.e., $\H=\sqrt{\beta}\widetilde{\bs{U}}$, where
\begin{equation}\label{eq:Ut}
    \widetilde{\bs{U}}=\bs{U}\begin{bmatrix}\mathbf{I}_K \\ \bs{0}_{(M-K)\times K}\end{bmatrix},
\end{equation}
and $\bs{U}\in \mathcal{U}(M)$ ($M \times M$ unitary). Note the slight abuse of notation so that $\sqrt{\beta}$ corresponds to the singular values of the orthogonal channel. We then have
\begin{equation}
    \H^\mrm{H} \H = \beta \mathbf{I}_K.
\end{equation}

Since the early research on MIMO systems, orthogonal channels were found to be desirable for several reasons \cite{mimo}:
\begin{itemize}
    \item Full multiplexing gain is available since all eigenvalues of the channel matrix are non-zero.
    \item Waterfilling algorithms are not required for maximizing capacity since all eigenvalues of the channel are equal.
    \item In the case of MU-MIMO, the users are served fairly since the different spatial streams have equal power.
    \item Simple linear equalization or precoding, namely MRC or MRT, achieves optimum performance, since it can exploit the orthogonal paths of the channel without the need for UE cooperation in MU-MIMO.
\end{itemize}

We next show how to construct $\bs{\Theta}$, for the case of \RISA{} and \RISAA, so that the resulting channel \eqref{eq:channel} is orthogonal.
\subsection{\RISA}
We are interested in finding $\alpha_1, \dots, \alpha_N$ such that
\begin{equation}\label{eq:RISA_eq}
    \H_0+\H_1 \bs{\Theta}_\mrm{\RISA} \H_2 = \sqrt{\beta}\widetilde{\bs{U}},
\end{equation}
Let us define
\begin{equation*}
    \H_1 = \begin{bmatrix} \bs{h}_{11},\dots, \bs{h}_{1N}\end{bmatrix},\;\;\;\; \H_2 = \begin{bmatrix} \bs{h}_{21},\dots, \bs{h}_{2N}\end{bmatrix}^\mrm{T},
\end{equation*}
where $\bs{h}_{1i}$ corresponds to column $i$ of $\H_1$ and $\bs{h}_{2i}^\mrm{T}$ corresponds to row $i$ of $\H_2$. We can then rewrite \eqref{eq:RISA_eq} as
\begin{equation}\label{eq:RISA_eq2}
    \sum_{i=1}^N \alpha_i \bs{h}_{1i}\bs{h}_{2i}^\mrm{T}=\sqrt{\beta}\widetilde{\bs{U}}-\H_0.
\end{equation}
By noting that \eqref{eq:RISA_eq2} is a linear equation in the vector $\bs{\alpha}=\begin{bmatrix}\alpha_1,\dots,\alpha_N\end{bmatrix}^\mrm{T}$, we can use the vectorization operation to reach
\begin{equation}\label{eq:RISA_eq3}
    \bs{\mathcal{H}}_{12}\bs{\alpha} = \mrm{vec}\left(\sqrt{\beta}\widetilde{\bs{U}}-\H_0\right),
\end{equation}
where $\bs{\mathcal{H}}_{12}=\begin{bmatrix}\vec(\bs{h}_{11}\bs{h}_{21}^\mrm{T}) & \dots & \vec(\bs{h}_{1N}\bs{h}_{2N}^\mrm{T})\end{bmatrix}$, which corresponds to an $MK \times N$ matrix. Assuming $\bs{\mathcal{H}}_{12}$ is full-rank, \eqref{eq:RISA_eq3} leads to an orthogonalization requirement for \RISA, namely $N\geq M K$. We would then solve \eqref{eq:RISA_eq3} by
\begin{equation}\label{eq:RISA_alph}
    \bs{\alpha} = \bs{\mathcal{H}}_{12}^\dagger\mrm{vec}\left(\sqrt{\beta}\widetilde{\bs{U}}-\H_0\right),
\end{equation}
where $\bs{\mathcal{H}}_{12}^\dagger$ is the right pseudo-inverse\footnote{Note that, although we can generate different right pseudo-inverses by adding matrices in the null-space of $\bs{\mathcal{H}}_{12}$, the common expression for right pseudo-inverse $\bs{\mathcal{H}}_{12}^\dagger=\bs{\mathcal{H}}_{12}^\mrm{H}(\bs{\mathcal{H}}_{12}\bs{\mathcal{H}}_{12}^\mrm{H})^{-1}$ minimizes the norm of $\bs{\alpha}$ for the given $\widetilde{\bs{U}}$ and $\beta$, which is most desirable in this work.} of $\bs{\mathcal{H}}_{12}$. Note that for obtaining $\bs{\alpha}$ we have not used the fact that the desired channel should be orthogonal. In fact, we could generate any channel matrix if we substitute $\sqrt{\beta}\widetilde{\bs{U}}$ in \eqref{eq:RISA_alph} by the desired channel.

\subsection{\RISAA}
We are  interested in finding a full-matrix $\bs{\Theta}_\mrm{\RISAA}$ such that
\begin{equation}\label{eq:RISAA_eq}
    \H_0+\H_1 \bs{\Theta}_\mrm{\RISAA} \H_2 = \sqrt{\beta}\widetilde{\bs{U}}.
\end{equation}
Assuming $\H_1$ and $\H_2$ are full-rank, we can select the reflection matrix as $\bs{\Theta}_\mrm{\RISAA}=\H_1^\dagger \bs{B} \H_2^\dagger$, where $\H_1^\dagger$ is the right pseudo-inverse of $\H_1$, $\H_2^\dagger$ is the left pseudo-inverse of $\H_2$, and $\bs{B}$ is an $M\times K$ matrix to be selected. This removes, with minimum power, the effect of $\H_1$ and $\H_2$ on the overall channel, and gives the orthogonalization requirement for \RISAA{} $N\geq \min(M,K)$, which, given $M>K$, leads to $N\geq M$. We then get
\begin{equation}\label{eq:RISAA_eq2}
    \bs{\Theta}_\mrm{\RISAA} = \H_1^\dagger \left( \sqrt{\beta}\widetilde{\bs{U}}-\H_0 \right)\H_2^\dagger.
\end{equation}
As happened with \RISA{}, we can also generate a non-orthogonal channel matrix by substituting $\sqrt{\beta}\widetilde{\bs{U}}$ in \eqref{eq:RISAA_eq2} with any other channel matrix.

\subsection{RIS baseline}
Achieving perfect channel orthogonalization is generally not possible if we consider the widely studied RIS model \eqref{eq:mod_RIS}. In case there existed a solution, it would come from finding a vector $\bs{\alpha}$ in \eqref{eq:RISA_alph} such that $\vert\alpha_n\vert^2=1$ $\forall n$. Obtaining said solution would correspond to finding a combination of  $\widetilde{\bs{U}}$ (from a subspace of the unitary matrices), $\beta$, and a vector in the null-space of $\bs{\mathcal{H}}_{12}$ leading to a solution of \eqref{eq:RISA_eq3} with  $\vert\alpha_n\vert^2=1$ $\forall n$. This problem seems analytically intractable, so we can only restrict ourselves to approximate solutions by numerical optimization. Since our goal is channel orthogonalization, we can find approximate solutions by numerical minimization of
\begin{equation}\label{eq:RIS_opt}
    \min_{\phi_1,\dots,\phi_N} \kappa(\H_0+\H_1\bs{\Theta}_\mrm{RIS}\H_2),
\end{equation}
where $\kappa(\cdot)$ is the  condition number of a matrix, given by the division between its maximum and minimum singular value, i.e., $\kappa(\cdot)=\sigma_{\max}(\cdot)/\sigma_{\min}(\cdot)$. Note that $\kappa(\cdot)\geq 1$, with equality only for orthogonal matrices. Thus, by minimizing it we would achieve a channel as close as possible to orthogonal, which will be used as a baseline approach.

\section{Channel estimation and RS configuration}\label{sec:ch}
In this section, we propose two techniques (one for \RISA{} and one for \RISAA{}) for estimating the channel and RS configuration at the BS. The idea is that, since it is desirable for RSs to have limited energy consumption\cite{RIS_vs_relay,RIS_survey}, and thus limited computation capabilities, we propose to leave most of the task of channel estimation and RS weight computation to the BS. For the sake of simplicity, we assume that each channel estimation step works perfectly. Proposing more specific channel estimation methods and characterizing the estimation errors is left as future work.
\subsection{\RISA{} configuration}
We start by proposing a method to configure the \RISA{} and obtain the desired channel. The main goal is to estimate the necessary channel information at the BS to be able to compute $\bs{\alpha}$ given by \eqref{eq:RISA_alph}. Since the channel matrix can be arbitrarily chosen by selecting $\sqrt{\beta}\widetilde{\bs{U}}$ in \eqref{eq:RISA_alph} (recall it needs not be orthogonal), we can assume that it is the BS itself that selects the desired channel so that it does not need to further estimate it. The following steps describe the method for finding the \RISA{} configuration at the BS:
\subsubsection{Estimation of $\H_0$} First, the \RISA{} fixes $\bs{\alpha}=\bs{0}_{N\times 1}$, and the UEs send $K$ orthogonal pilots. The received symbols over $K$ slots would be then given by the $M\times K$ matrix
\begin{equation}\label{eq:RISAest_Y1}
    \bs{Y}_1 = \H_0\bs{P}+ \bs{N}_1,
\end{equation}
where $\bs{P}$ is the previously known pilot matrix, which can be fixed to, e.g., $\bs{P}=\mathbf{I}_K$, and $\bs{N}_1$ is the noise matrix with IID entries $n_{ij}\sim \mathcal{CN}(0, N_0)$. From \eqref{eq:RISAest_Y1} we can directly estimate $\H_0$ using state-of-the art channel estimation methods.

\subsubsection{Estimation of $\bs{\mathcal{H}}_{12}$} Since the effect of $\H_1$ and $\H_2$ for the selection of $\bs{\alpha}$ only comes through $\bs{\mathcal{H}}_{12}$, it is enough to estimate said matrix, whose columns are given by
\begin{equation}\label{eq:col_H12}
    [\bs{\mathcal{H}}_{12}]_{:,n} = \vec(\bs{h}_{1n}\bs{h}_{2n}^\mrm{T}).
\end{equation}
Let us assume that the \RISA{} is configured such that for a given $n$ we have $\alpha_n=1$ and $\alpha_i=0$ for $i\neq n$. Transmitting $K$ orthogonal pilots from the UEs would then lead to
\begin{equation}
    \bs{Y}_{2n} = \H_0 \bs{P}+ \bs{h}_{1n}\bs{h}_{2n}^\mrm{T}\bs{P}+\bs{N}_{2n},
\end{equation}
where $\bs{P}$ and $\bs{N}_{2n}$ are defined as before.
Assuming we already know $\H_0$ from the previous step, we can cancel it at the BS, leading to
\begin{equation}\label{eq:RISAest_Y2}
    \widetilde{\bs{Y}}_{2n} =\bs{h}_{1n}\bs{h}_{2n}^\mrm{T}\bs{P}+\widetilde{\bs{N}}_{2n},
\end{equation}
where $\widetilde{\bs{N}}_{2n}$ would include the estimation error from the previous step, which could lead to correlated entries. From \eqref{eq:RISAest_Y2} we can estimate $\bs{h}_{1n}\bs{h}_{2n}^\mrm{T}$ using state-of-the-art estimation methods, which, after vectorizing, would give the estimate of the $n$th column of $\bs{\mathcal{H}}_{12}$ given in \eqref{eq:col_H12}. If we iteratively perform this estimation step for $n=1,\dots ,N$, i.e., setting to $1$ each element of the \RISA{} at a time, the BS would construct a full estimate of $\bs{\mathcal{H}}_{12}$.
\subsubsection{Computation and forwarding of $\bs{\alpha}$} Once the BS has estimated $\H_0$ and $\bs{\mathcal{H}}_{12}$ it can select its desired channel, which in our case corresponds to $\sqrt{\beta}\widetilde{\bs{U}}$, and compute $\bs{\alpha}$ using \eqref{eq:RISA_alph}. The BS should then forward $\bs{\alpha}$ to the \RISA{} so that it can be configured to create the desired channel, which is already known at the BS.

The proposed method allows to configure the \RISA{} to generate any channel matrix by using a total of $(N+1)K$ pilot slots. This corresponds to a notable decrease with respect to estimating $\H_0$, $\H_1$, and $\H_2$ independently, which would at least require $MK+N(M+K)$ slots, or even more for practical methods such as in \cite{IRS_joham} for RIS. Moreover, since the BS selects its desired channel, it can directly use it for equalization/precoding purposes. Also, form the restriction of the channels to be orthogonal, optimum equalization/precoding would be achieved through simple MRC/MRT, i.e., multiplying the conjugate transpose of the channel matrix at the BS.

\subsection{\RISAA{} configuration}
If we inspect \eqref{eq:RISAA_eq2}, we note that we need an estimate of both $\H_1$ and $\H_2$ to be able to compute the corresponding \RISAA{} configuration, so an efficient method such as the the one for \RISA{} may not be available. Let us thus consider that the \RISAA{} can transmit pilots through each of its elements. Although this might not be desirable in practice, the concept of \RISAA{} is not yet well-established in contemporary literature, so we  use this assumption as a first step towards defining the operation of such RSs. Coming up with more elaborate methods to avoid the requirement of sending pilots from the \RISAA{} will be considered in future work. The following steps describe the proposed method for \RISAA{} configuration:
\subsubsection{Estimation of $\H_0$}
In the initial step the \RISAA{} would fix $\bs{\Theta}_{\RISA}=\bs{0}_{N\times N}$, and the UEs would send pilots to perform the channel estimation of $\H_0$ as in the \RISA{} case.

\subsubsection{Estimation of $\H_1$}
In the case of \RISAA{} we need to have an estimate of $\H_1$ and $\H_2$ to compute their pseudo-inverses in \eqref{eq:RISAA_eq2}. With the assumption that the \RISAA{} has the ability to send pilots through each of its elements, the \RISAA{} would send $N$ orthogonal pilots leading to the received matrix at the BS
\begin{equation}\label{eq:FRIS_H1est}
    \bs{Y}_2 = \H_1 \bs{P}_\mrm{\RISAA{}}+\bs{N}_2,
\end{equation}
where $\bs{P}_\mrm{\RISAA{}}$ is the $N\times N$ known pilot matrix, which can be set to $\mathbf{I}_N$. From \eqref{eq:FRIS_H1est}, the BS can estimate $\H_1$ using state-of-the-art estimation methods.

\subsubsection{Estimation of $\H_2$}
Let the \RISAA{} fix alternatively each group of $M$ elements to 1, i.e., at instant $n$ we select
\begin{equation}
    \bs{\Theta}_\mrm{\RISAA}=\diag(\begin{bmatrix}\bs{0}_{1\times (n-1)M} & \bs{1}_{1\times M} & \bs{0}_{1\times (N-nM)}\end{bmatrix}),
\end{equation}
and we send $K$ orthogonal pilots from the UEs. The BS would then receive
\begin{equation}\label{eq:FRIS_H1est1}
    \bs{Y}_{3n} = \H_0\bs{P}+\H_{1,\mrm{sq}}(n)\H_{2,\mrm{sq}}(n) \bs{P}+\bs{N}_3,
\end{equation}
where $\H_{1,sq}(n)$ is the $M\times M$ matrix formed by the columns $(n-1)M+1$ to $n M$ of $\H_1$, $\H_{2,sq}(n)$ is the $M\times K$ matrix formed by rows $(n-1)M+1$ to $n M$ of $\H_2$, and $\bs{N}_3$ and $\bs{P}$ are the noise and pilot matrix, respectively. Assuming $\H_{1,sq}(n)$ is full-rank $\forall n$, we can get an estimate of $\H_{2,sq}(n)$ applying state-of-the-art estimation methods to
\begin{equation}\label{eq:FRIS_H1est2}
    \widetilde{\bs{Y}}_{3n} = \widehat{\H}_{1,\mrm{sq}}^{-1}(n)(\H_{1,\mrm{sq}}(n)\H_{2,\mrm{sq}}(n) \bs{P}+\bs{N}_3),
\end{equation}
where $\widehat{\H}_{1,\mrm{sq}}(n)$ is the estimate of $\H_{1,\mrm{sq}}(n)$  from the previous stage. In the last instant, given by $n=\left\lceil N/M \right\rceil,$ $nM$ might exceed $N$, so everything should be cropped to $N$ in \eqref{eq:FRIS_H1est2}, and we would change the inverse for the left pseudo-inverse of the cropped $\widehat{\H}_1$. Note that the assumption of having $\H_{1,\mrm{sq}}(n)$ of rank $M$ $\forall n$ is a bit more restrictive than the requirement of solvability of \eqref{eq:RISAA_eq}, where only the whole matrix $\H_1$ should be rank $M$. However, in case some $\H_{1,\mrm{sq}}(n)$ are ill-conditioned, which can be known at the BS from the estimate of $\H_1$, we could think of alternative solutions, e.g., selecting groups of $M$ linearly independent rows. In the worst case, we could also fix a smaller number of 1s in the \RISAA{} and use the pseudo-inverse instead of inverse of the resulting cropped $\H_1$, but this would require larger number of pilot slots.

\subsubsection{Computation and forwarding of $\bs{\Theta}_\mrm{\RISAA}$} As a final step, the BS would select the desired channel ($\sqrt{\beta}\widetilde{\bs{U}}$) and compute the \RISAA{} configuration, $\bs{\Theta}_\mrm{\RISAA}$, using \eqref{eq:RISAA_eq2} with the estimates of $\H_0$, $\H_1$, and $\H_2$. The BS would then forward $\bs{\Theta}_\mrm{\RISAA}$ to the \RISAA{}, which would then apply it.

The proposed method allows the BS to configure the \RISAA{} for inducing some desired channel, in this case orthogonal, by employing a total of $\left(1+\left\lceil N/M \right\rceil \right)K+N$ pilots, where $N$ of them would correspond to pilots sent from the \RISAA{}.  For a moderate number of users, this leads to a notable decrease with respect to the \RISA{} method, which requires $(N+1)K$ pilots. Furthermore, we should note that the required $N$ for \RISAA{} can also be remarkably smaller than for \RISA{}. A summary of the orthogonalization conditions for each RS can be found in Table~\ref{tab:comp_RS}.

\section{RS power constraints}\label{sec:min}
In this section we study the problem of reducing the power requirements for the RS configurations achieving channel orthogonality. As shown in Table~\ref{tab:comp_RS}, we define the power of the different RS settings as the squared Frobenius norm of the reflection matrix $\bs{\Theta}$, which corresponds to the sum power throughout its entries. Let us then assume that each RS can operate without amplification as long as the average power per RS element is no greater than 1 (RIS achieves this with equality), which translates to $\Vert\bs{\Theta}\Vert^2_\mrm{fro}\leq N$. Note that, ideally, each RS element should have power no greater than 1, which will be considered in the extended version of the paper.

Another factor to consider is the power of the resulting orthogonal sub-channels of $\H=\sqrt{\beta}\widetilde{\bs{U}}$. Said power, given by $\beta$ (orthogonal channels have all eigenvalues equal), would be linearly related to the post-processed SNR per UE (after MRC/MRT),
$\eta = \beta E_s/N_0$,
where we have assumed that the RS does not introduce extra noise.\footnote{RSs with amplification might suffer from noise enhancement similar to that of zero-forzing (ZF) equalizers. A thorough characterization of it may be considered in future work.} Recall that, from the orthogonality of the channel, there is no interference between UEs and all UEs have the same post-processed SNR. Thus, for a limited RS power, we would ideally like to have a large $\beta$ so as to increase the capacity per UE .
\subsection{\RISA}
The \RISA{} sum power  required for having $\H=\sqrt{\beta}\widetilde{\bs{U}}$ is given by (see Table~\ref{tab:comp_RS})
\begin{equation}\label{eq:Paris}
    P_\mrm{A}(\beta,\widetilde{\bs{U}}) = \beta g_1(\widetilde{\bs{U}})-2\sqrt{\beta}f_1(\widetilde{\bs{U}})+c_1,
\end{equation}
where we defined $f_1(\widetilde{\bs{U}})\!=\!\mathfrak{Re}\left\{ \vec(\widetilde{\bs{U}})^\mrm{H} \bs{G}_{12}^{-1} \vec(\H_0)\right\}$, $g_1(\widetilde{\bs{U}})\!=\!\vec(\widetilde{\bs{U}})^\mrm{H}\bs{G}_{12}^{-1}\vec(\widetilde{\bs{U}})$, $c_1\!=\!\vec(\H_0)^\mrm{H}\bs{G}_{12}^{-1}\vec(\H_0)$,  with $\bs{G}_{12}=\bs{\mathcal{H}}_{12}\bs{\mathcal{H}}_{12}^\mrm{H}$. Equation \eqref{eq:Paris} comes from substituting \eqref{eq:RISA_alph} in the \RISA{} power expression from Table~\ref{tab:comp_RS} and operating. Let us first focus on obtaining the minimum \RISA{} power for achieving an orthogonal channel. We can immediately note that the existence of the direct channel $\H_0$ is responsible for requiring a minimum power to be able to orthogonalize the channel with \RISA{}. In the absence of $\H_0$ ($c_1=f_1(\widetilde{\bs{U}})=0$), $P_\mrm{A}(\beta,\widetilde{\bs{U}})$ can be made arbitrarily small by lowering $\beta$, i.e., sacrificing SNR; therefore, channel orthogonalization would be achievable without the need for amplification. Let us then assume $\H_0$ is present. Note that the BS has freedom in selecting $\widetilde{\bs{U}}$ and $\beta$. We can then obtain the minimum power required for orthogonalization with \RISA{} by solving
\begin{equation}\label{eq:opt_ARIS}
\begin{aligned}
    P_\mrm{A,min}=\;\;&\min_{\beta, \widetilde{\bs{U}}} \;\; P_\mrm{A}(\beta, \widetilde{\bs{U}})\\
    & \;\; \mrm{s.t. } \;\; \widetilde{\bs{U}}^\mrm{H}\widetilde{\bs{U}}=\mathbf{I}_K.
\end{aligned}
\end{equation}
Differentiating $P_\mrm{A}(\beta, \widetilde{\bs{U}})$ over $\beta$ and equalling to 0 gives us the minimum $\beta$
\begin{equation}
    \beta_{\mrm{o}1} = \left(\frac{f_1(\widetilde{\bs{U}})}{g_1(\widetilde{\bs{U}})}\right)^2.
\end{equation}
We can then substitute $\beta_{\mrm{o}1}$ in \eqref{eq:Paris} to get $P_\mrm{A}(\beta_{\mrm{o}1}, \widetilde{\bs{U}})$, which can then be minimized using gradient descent within the unitary space. In order to improve accuracy of the optimization, we consider optimization over the geodesics of the unitary space as proposed in \cite{traian}. Thus, we need to obtain the Euclidean gradient by differentiating $P_\mrm{A}(\beta_*, \widetilde{\bs{U}})$ over $\bs{U}^*$ (recall \eqref{eq:Ut}), and use it for algorithm in \cite[Table~II]{traian}, which includes Armijo line-search for better convergence. We get
\begin{equation}\label{eq:diff_Paris}
\begin{aligned}
    \frac{\partial P_\mrm{A}(\beta_{\mrm{o}1}, \widetilde{\bs{U}})}{\partial \widetilde{\bs{U}}^*} = \frac{b}{g_1^2(\widetilde{\bs{U}})}\vec^{-1}\Big(-f_1^2(\widetilde{\bs{U}})\bs{G}_{12}^{-1} \vec(\widetilde{\bs{U}}) &
    \\+f_1(\widetilde{\bs{U}})g_1(\widetilde{\bs{U}})\bs{G}_{12}^{-1}\vec(\H_0) \Big)&,
\end{aligned}
\end{equation}
where $b=1-2 \mrm{sign}\left(f_1(\widetilde{\bs{U}})\right)$. Note that, for differentiating over $\bs{U}^*$ instead of $\widetilde{\bs{U}}^*$, we would just complete \eqref{eq:diff_Paris} with zeros, since the corresponding extra columns of $\bs{U}$ have no bearing on $P_\mrm{A}(\beta_{\mrm{o}1}, \widetilde{\bs{U}})$. Once we have obtained $P_\mrm{A,min}$, any other ARIS sum power above it can be achieved from \eqref{eq:Paris} by solving a second order equation over $\sqrt{\beta}$. Note that for every different $\beta$ there may be a new optimal $\widetilde{\bs{U}}$, i.e., different from the one solving \eqref{eq:opt_ARIS}, which minimizes the resulting power. Alternatives of \eqref{eq:opt_ARIS} will be studied in the extended version.
\subsection{\RISAA}
The \RISAA{} sum power giving $\H=\sqrt{\beta}\widetilde{\bs{U}}$ corresponds to
\begin{equation}\label{eq:Pfris}
\begin{aligned}
    P_\mrm{F} = & \beta g_2(\widetilde{\bs{U}})-2\sqrt{\beta}f_2(\widetilde{\bs{U}})+c_2,
\end{aligned}
\end{equation}
where we defined $f_2(\widetilde{\bs{U}})\!=\!\mathfrak{Re}\left\{\mrm{tr}(\bs{G}_2^{-1}\widetilde{\bs{U}}^\mrm{H}\bs{G}_1^{-1}\H_0)\right\}$, $g_2(\widetilde{\bs{U}})\!=\!\mrm{tr}(\bs{G}_2^{-1}\widetilde{\bs{U}}^\mrm{H}\bs{G}_1^{-1}\widetilde{\bs{U}})$, $c_2\!=\!\mrm{tr}(\bs{G}_2^{-1}\H_0^\mrm{H}\bs{G}_1^{-1}\H_0)$, with $\bs{G}_1=\H_1\H_1^\mrm{H}$ and $\bs{G}_2=\H_2^\mrm{H}\H_2$. We can use the same reasoning as in the case for \RISA{} throughout the different steps. Let us thus focus on solving
\begin{equation}\label{eq:opt_Pfris}
\begin{aligned}
   P_\mrm{F,min} =\;\; &\min_{\beta, \widetilde{\bs{U}}} \;\; P_\mrm{F}(\beta, \widetilde{\bs{U}})\\
    & \;\; \mrm{s.t. } \;\; \widetilde{\bs{U}}^\mrm{H}\widetilde{\bs{U}}=\mathbf{I}_K.
\end{aligned}
\end{equation}
Proceeding as in the previous case we can get
\begin{equation}
    \beta_{\mrm{o}2} = \left(\frac{f_2(\widetilde{\bs{U}})}{g_2(\widetilde{\bs{U}})}\right)^2,
\end{equation}
which leads to the euclidean gradient to be used for minimizing over $\widetilde{\bs{U}}$ using \cite[Table~II]{traian},
\begin{equation}\label{eq:diff_Pfris}
\begin{aligned}
    \frac{\partial P_\mrm{F}(\beta_{\mrm{o}1}, \widetilde{\bs{U}})}{\partial \widetilde{\bs{U}}^*} = \frac{b}{g_2^2(\widetilde{\bs{U}})}\!\Big(\!-f_2^2(\widetilde{\bs{U}})\bs{G}_{1}^{-1} \widetilde{\bs{U}}^\mrm{H} \bs{G}_{2}^{-1}&
    \\+g_2(\widetilde{\bs{U}})f_2(\widetilde{\bs{U}})\bs{G}_{1}^{-1}\H_0\bs{G}_{2}^{-1} \Big)&,
\end{aligned}
\end{equation}
where $b=1-2 \mrm{sign}\left(f_2(\widetilde{\bs{U}})\right)$.

\begin{table*}[]
    \renewcommand{\arraystretch}{1.6}
    \centering
    \begin{tabular}{|c||c|c|c|}
    \hline
         & \textbf{\RISA{}} & \textbf{\RISAA{}} & \textbf{RIS} \\
         \hline \hline
         \textbf{Minimum $N$ for orthogonalization} & $MK$ & $\min(M,K)$ & -\\
         \hline
         \textbf{Number of pilots} & $(N+1)K$ & $\left(1+\left\lceil \frac{N}{M} \right\rceil \right)K+N$ & $>MK+N(M+K)$ \cite{IRS_joham}\\
         \hline
         \textbf{RS sum power} & $\Vert \Theta_\mrm{\RISA{}}\Vert^2_{\mrm{fro}}=\bs{\alpha}^\mrm{H}\bs{\alpha}$ & $\Vert \Theta_\mrm{\RISAA{}}\Vert^2_{\mrm{fro}}=\mrm{tr}(\Theta_\mrm{\RISAA{}}^\mrm{
         H}\Theta_\mrm{\RISAA{}})$ & $\Vert\Theta_\mrm{RIS}\Vert^2_\mrm{fro}=N$ \\
         \hline
    \end{tabular}
    \caption{Orthogonalization conditions for different RSs.}
    \label{tab:comp_RS}
\end{table*}

\section{Numerical results}\label{section:num_res}
For the numerical results, we have tried to solve the optimization problems defined in \eqref{eq:opt_ARIS} and \eqref{eq:opt_Pfris}. Finding closed form results for said problems is in general intractable due the constraint in $\widetilde{\bs{U}}$, which should live in a subspace of the unitary matrices. However, good local solutions can be found by using gradient descent along the geodesics, as proposed in \cite{traian}. We cannot assure that the obtained results reach absolute minima, but, since our main goal is to check if the proposed RS technologies can be realized without amplification, local minima may be enough for our purpose. We have thus implemented \cite[Table~II]{traian} with the Euclidean gradients defined in \eqref{eq:diff_Paris} and \eqref{eq:diff_Pfris} to find the minimum power required for perfect channel orthongonalization using ARIS and FRIS, respectively.

\begin{figure}
     \centering
     \begin{subfigure}[b]{0.24\textwidth}
         \centering
         \includegraphics[scale=0.25]{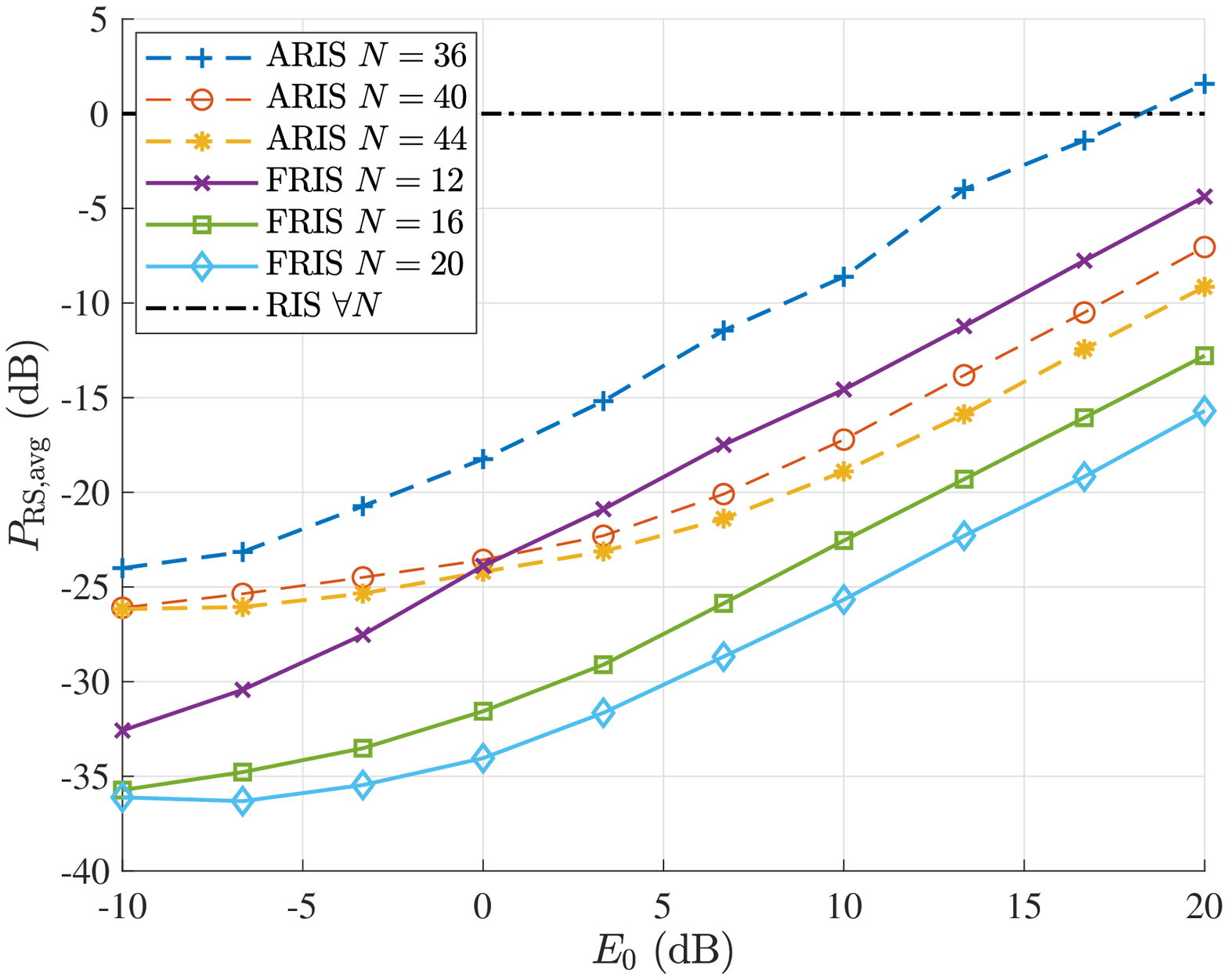}
     \end{subfigure}
     \hfill
     \begin{subfigure}[b]{0.237\textwidth}
         \centering
         \includegraphics[scale=0.25]{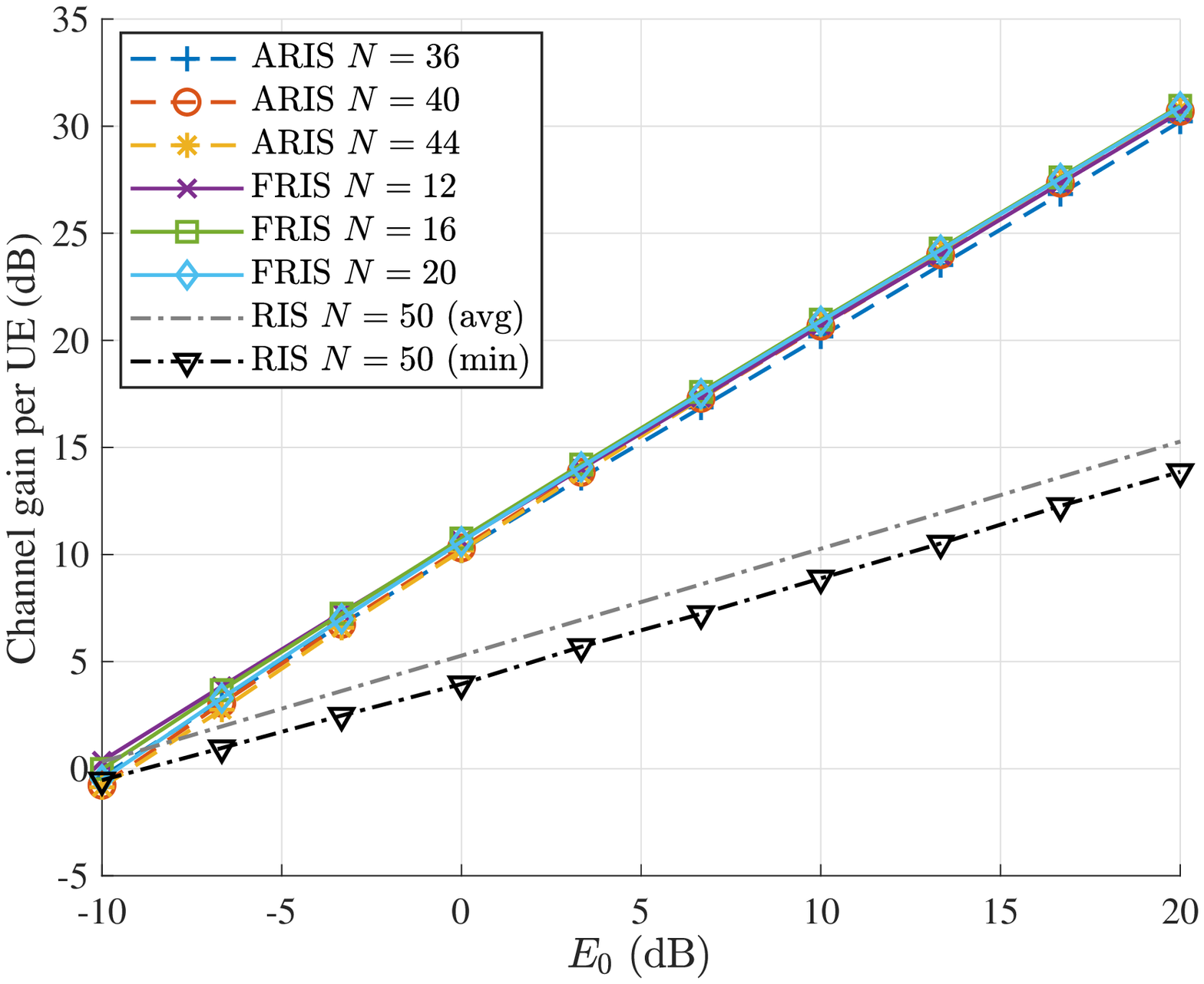}
     \end{subfigure}
     \caption{Minimum average RS power per element (left) and resulting minimum channel gain per UE with respect to normalized gain of the direct channel.}
        \label{fig:Pmin}
\end{figure}

\begin{figure}
     \centering
     \includegraphics[scale=0.3]{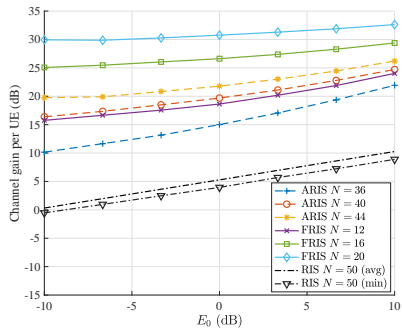}
        \caption{Channel gain per UE for unit average RS power per element with respect to normalized gain of the direct channel.}
        \label{fig:ChGain}
\end{figure}

In Fig.~\ref{fig:Pmin} we can see the minimum average RS power per element, $P_\mrm{RS,avg}=P_{\{\mrm{A,F}\}\mrm{,min}}/N$, and the resulting channel gain per UE, equal to $\beta$ for all UEs from the orthogonal restriction, with respect to the normalized power of the direct channel, $E_0$. Since we are most interested in the power relation between the direct and reflected channels, we have used normalized IID Rayleigh fading channels with $\Vert H_0 \Vert^2_{\mrm{Fro}}=E_0MK$, $\Vert H_1 \Vert^2_{\mrm{Fro}}=MN$, $\Vert H_2 \Vert^2_{\mrm{Fro}}=NK$. Other channel models will be considered in future work, but we may note from the analytical results that ill-conditioned channels are most harmful in the RS-reflected paths. Fig.~\ref{fig:Pmin} (left) shows that in most practical scenarios (direct links with power below 100 times the reflected one), the minimum average power for channel orthogonalization, with both ARIS and FRIS, can be smaller than that of RIS, so these surfaces could potentially be implemented without amplification. The resulting channel gains for these minimized powers have analogous linear relation with $E_0$, still impressive since they even outperform RIS, which has been numerically optimized for channel orthogonalization using \eqref{eq:RIS_opt}.\footnote{For the RIS, since perfect orthogonality may not be reachable, we plotted the average channel gain and minimum channel gain per UE.} However, the results for RIS may be far from optimum due to the difficulty of such task, and the analytical intractability. Finding more suitable optimization formulations for channel orthogonalization with RIS should be further studied. A important thing to note is that the channel gains in Fig.~\ref{fig:Pmin} are achieved with RS power dependent on $E_0$, and generally below that of RIS. If we increase the respective gains until all RS powers are equal to that of RIS, the resulting channel gains, which are plotted in Fig.~\ref{fig:ChGain}, are even more impressive, especially for FRIS, which can get 10 times better channel gains than ARIS with a lower number of elements. In fact, there is room for improvement by further optimization of $\widetilde{\bs{U}}$, as previously discussed.

\section{Conclusions}\label{section:conc}
We have presented the concepts of FRIS and ARIS, two alternative RS technologies with relaxed restrictions over RIS. We have obtained analytical results for FRIS and ARIS configurations that achieve perfect channel orthogonalization. We proposed a channel estimation method for each RS technology at the BS, which selects the desired channel and forwards the corresponding RS configuration. We have also showed that these RS can perform channel orthogonalization without the need of amplification by minimizing over the unitary space. The achieved channel gains, which are fairly distributed among users from the orthogonalization, remark the benefits of adding more processing capabilities at the RSs.
\renewcommand{\baselinestretch}{.95}
\bibliographystyle{IEEEtran}
\bibliography{IEEEabrv,bibliography}

%

\end{document}